\documentclass{article}
\usepackage{ijcai25}

\usepackage{times}
\usepackage{soul}
\usepackage{url}
\usepackage[hidelinks]{hyperref}
\usepackage[utf8]{inputenc}
\usepackage[small]{caption}
\usepackage{multirow}
\usepackage{graphicx}
\usepackage{amsmath}
\usepackage{amsthm}
\usepackage{booktabs}
\usepackage{algorithm}
\usepackage{algorithmic}
\usepackage[switch]{lineno}
\usepackage{color}

\usepackage{float}
\usepackage{stfloats}
\usepackage{subfig}

\usepackage{amsmath}
\usepackage{amssymb}

\title{GSDNet: Revisiting Incomplete Multimodal-Diffusion from Graph Spectrum Perspective for Conversation Emotion Recognition}

\author{
Yuntao Shou$^{1}$
\and
Jun Yao$^2$\and
Tao Meng$^1$\and
Wei Ai$^1$ \and
Cen Chen$^3$\footnote{Corresponding author} \and
Keqin Li$^4$
\affiliations
$^1$College of Computer and Mathematics, Central South University of Forestry and Technology, Changsha, Hunan 410004, China\\
$^2$Department of Computer Science, Anhui Normal University, Anhui 24100, China\\
$^3$Future Technology Institute, South China University of Technology, Guangdong 510641, China\\
$^4$Department of Computer Science, State University of New York, New Paltz, New York 12561, USA\\
\emails
shouyuntao@stu.xjtu.edu.cn,
yj@ahnu.edu.cn,
mengtao@hnu.edu.cn,
aiwei@hnu.edu.cn,
chencen@scut.edu.cn,
lik@newpaltz.edu
}

\begin{document}

\maketitle

\begin{abstract}
Multimodal emotion recognition in conversations (MERC) aims to infer the speaker's emotional state by analyzing utterance information from multiple sources (i.e., video, audio, and text). Compared with unimodality, a more robust utterance representation can be obtained by fusing complementary semantic information from different modalities. However, the modality missing problem severely limits the performance of MERC in practical scenarios. Recent work has achieved impressive performance on modality completion using graph neural networks and diffusion models, respectively. This inspires us to combine these two dimensions through the graph diffusion model to obtain more powerful modal recovery capabilities. Unfortunately, existing graph diffusion models may destroy the connectivity and local structure of the graph by directly adding Gaussian noise to the adjacency matrix, resulting in the generated graph data being unable to retain the semantic and topological information of the original graph. To this end, we propose a novel Graph Spectral Diffusion Network (GSDNet), which maps Gaussian noise to the graph spectral space of missing modalities and recovers the missing data according to its original distribution. Compared with previous graph diffusion methods, GSDNet only affects the eigenvalues of the adjacency matrix instead of destroying the adjacency matrix directly, which can maintain the global topological information and important spectral features during the diffusion process. Extensive experiments have demonstrated that GSDNet achieves state-of-the-art emotion recognition performance in various modality loss scenarios.
\end{abstract}

\section{INTRODUCTION}
Multimodal emotion recognition in conversations (MERC) aims to build an emotion recognition model with cross-domain understanding, reasoning, and learning capabilities by integrating video, audio, and text data \cite{li2023decoupled}, \cite{tsai2019multimodal}. MERC research mainly focuses on how to effectively encode discriminative representations from different modalities and achieve more accurate information fusion and analysis \cite{ramesh2021zero}, \cite{ding2023lggnet}, \cite{shou2022conversational}, \cite{shou2023comprehensive}, \cite{meng2024deep}.

However, the modality missing problem is unavoidable in the real world, and it may severely degrade the performance of multimodal understanding models \cite{wang2023distribution}, \cite{meng2024multi}, \cite{shou2024adversarial}, \cite{ai2024gcn}. For example, the text modality may fail due to environmental noise interference, resulting in the inability to obtain valid text information. The acoustic modality may lose part of the sound information due to sensor failure \cite{meng2024masked}, \cite{shou2025contrastive}. The visual modality may be affected by factors such as poor lighting conditions, object occlusion, or privacy protection requirements, resulting in the inability to obtain image or video data \cite{shou2025dynamic}, \cite{ai2025revisiting}. Therefore, how to design and optimize a multimodal emotion recognition model for conversation that can cope with modality loss has become an important direction of current research \cite{wang2024incomplete}, \cite{zhang2024learning}.

\begin{figure}
	\centering
    \setlength{\abovecaptionskip}{0.cm}
	\includegraphics[width=1\linewidth]{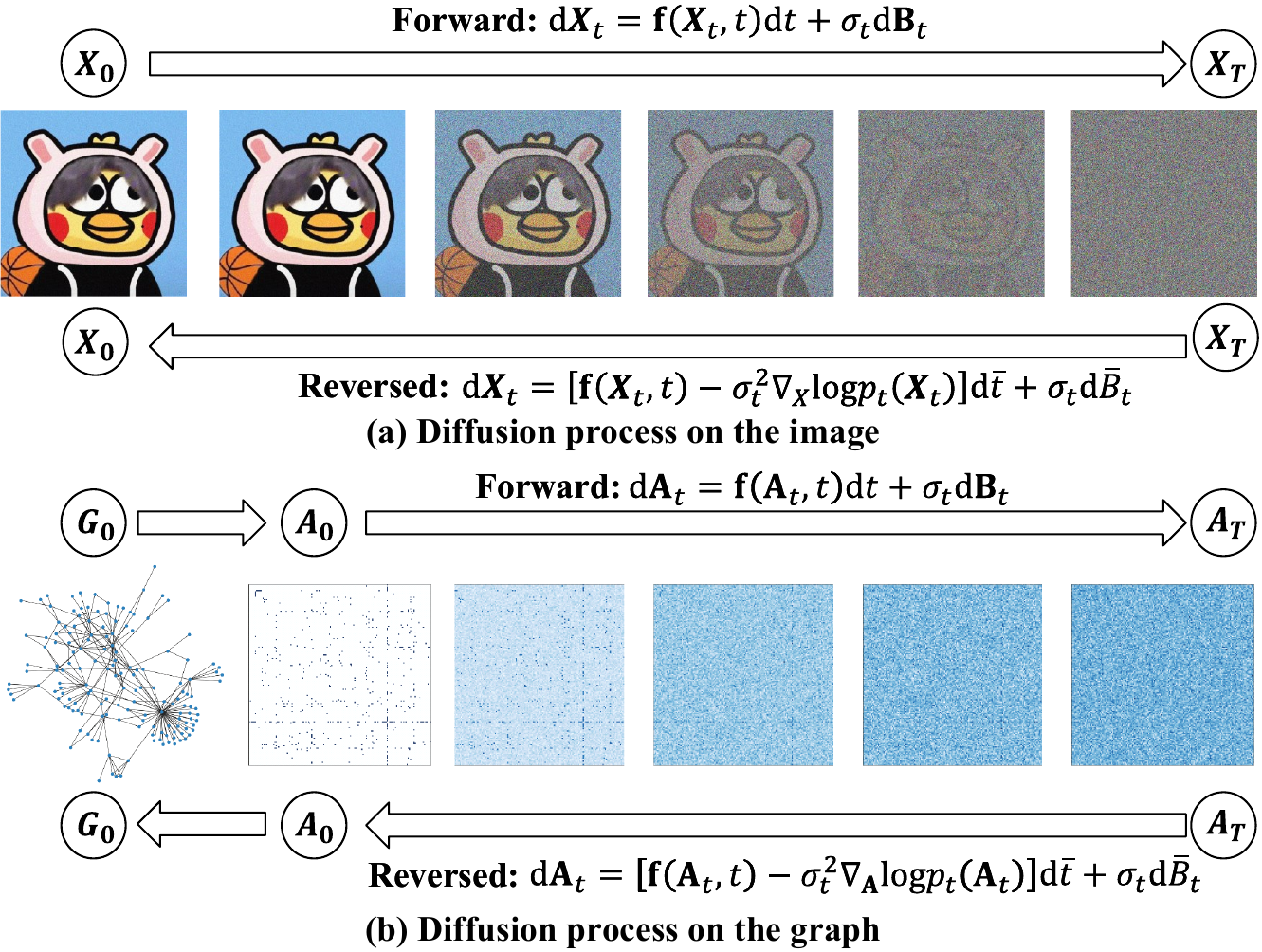}
	\caption{Illustration of the difference between images and graphs in the diffusion processes.}
	\label{fig:figure1}
    \vspace{-4mm}
\end{figure}

Recently, graph neural networks (GCNs) \cite{ai2023two}, \cite{shou2024low}, \cite{shou2024revisiting} and diffusion generative models (DGMs) have shown outstanding performance in multiple tasks such as vision, and language \cite{hu2021mmgcn}, \cite{jo2022score}. The core of GCNs is message passing based on graph structure to establish complex dependencies \cite{shou2024efficient}, \cite{shou2024spegcl}. The core of DGMs is forward denoising and backward denoising to learn the original distribution of data. Considering the potential of GCNs and DGMs, some researchers have tried to apply GCNs and SGMs to modality completion. They alleviate the impact of modality loss on multimodal emotion recognition performance from different perspectives. For GCNs, \cite{lian2023gcnet} applied graph completion networks to model semantic dependencies between different modal data to achieve modal recovery. For DGMs, \cite{wang2024incomplete} proposed modal diffusion to learn the original distribution of data to achieve modal recovery. All of the above methods have shown significant results because they have strong modeling capabilities in their respective dimensions. Specifically, GCNs and DGMs respectively model the dependence and distribution between multi-modal data to achieve modal recovery. Generally, fusing complex semantic information between multi-modal features and capturing the original distribution of multi-modal data are crucial for emotion recognition performance in missing modalities. This inspires us to combine these two dimensions to obtain more powerful modal recovery capabilities.

However, unlike visual data, which has dense structural information, the structure of graph data is generally sparse, causing the data generated by the graph diffusion model to be unable to retain the topological information of the original graph. As shown in Fig. \ref{fig:figure1}, the image perturbed by Gaussian noise still retains recognizable numerical patterns and local structural information in the early and middle stages of forward diffusion. For example, even if the details of the image are gradually covered by noise, its global contour and edge information can still be captured by the model to a certain extent. This enables the model to effectively restore the content of the image using this residual information during the reverse diffusion process. However, during the forward diffusion of graph data, the topological structure of the graph adjacency matrix is rapidly lost and a dense noise matrix is formed. Intuitively, the diffusion method of inserting Gaussian noise into the graph adjacency matrix seriously undermines the ability to learn the graph topology and feature representation. From a theoretical perspective, running diffusion over the entire space of the adjacency matrix will cause the signal-to-noise ratio (SNR) to drop rapidly and approach zero. Since the SNR is basically zero, the scoring network will not be able to effectively capture the gradient information of the original distribution during training.

To overcome these problems, we propose a novel Graph Spectral Diffusion Network (GSDNet) to strictly restrict the diffusion of Gaussian noise to the spectral space of the adjacency matrix. Specifically, we perform eigendecomposition on the adjacency matrix, decomposing it into eigenvalue matrix and eigenvector matrix and adding Gaussian noise to the eigenvalues without interfering with the eigenvectors. On the one hand, by operating the eigenvalues in the spectral space, the direct destruction of the local structure of the adjacency matrix can be effectively avoided, ensuring the generated graph still conforms to the global semantics. On the other hand, this constrained diffusion process can more naturally capture and preserve the topological information of the graph, ensuring that the generated graph has consistent spectral features. Overall, our contributions are as:

 \begin{itemize}
    \item We design a novel modality completion model, the Graph Spectral Diffusion Network (GSDNet), which can simultaneously model the dependencies between multimodal features and the distribution of original data to obtain powerful modality recovery capabilities.

    \item We strictly limit the diffusion of Gaussian noise in the spectral space of the adjacency matrix to avoid the destruction of the graph structure and ensure that the graph data generated by GSDNet still conforms to the global semantics.
 
     \item We conduct extensive experiments on multiple real-world datasets to demonstrate that our GSDNet outperforms state-of-the-art methods for conversational emotion recognition in incomplete multimodal scenarios.
 \end{itemize}

\section{Related Work}
\subsection{Incomplete Multimodal Learning}
In practical applications, missing modalities are an inevitable problem \cite{fu2025sdr}. To address this challenge, an effective approach is to find a low-dimensional subspace that can be shared by all modalities, in which the correlation between different modalities is maximized. However, strategies based on shared low-dimensional subspaces often ignore the complementarity between heterogeneous modalities. To overcome this shortcoming, another more effective approach is to restore the missing modality through the existing modality. This process not only requires inferring the content of the missing modality based on the features of the known modality but also ensures that the restored modality can work together with other modalities. Existing modality restoration methods can be divided into several types, including zero-based restoration \cite{parthasarathy2020training}, average-based restoration \cite{zhang2020deep}, and deep learning-based restoration \cite{pham2019found}. Since zero-filling and average-based restoration methods do not use any supervised information, the data they restore often have a significant gap with the original data. In contrast, deep learning-based methods, with their powerful feature learning capabilities, can more accurately estimate the missing modality. For example, Tran et al. \cite{tran2017missing} used a cascaded residual autoencoder to restore the missing modality, and the network's residual learning mechanism made the restoration effect more accurate. In addition, some researchers have proposed deep learning methods based on cross-modal restoration strategies, using cycle consistency loss to ensure the matching degree between the restored modality and the original modality \cite{zhao2021missing}. Other studies use graph neural networks (GNNs) to solve the modality restoration problem. For example, Lian et al. \cite{lian2023gcnet} introduced a graph neural network framework and combined the relationship between nodes and edges to enhance the correlation between modalities.

\subsection{Score-based Generative Models}
Score-based generative models (SGMs) estimate the probability distribution of data by parameterizing the score function \cite{song2019generative}, \cite{song2020improved}, \cite{songscore}. Specifically, SGMs model the scoring network $s(x;\theta)$ through learnable parameters $\theta$, thereby training the model to estimate $\nabla_{x}\log p(x)$. Unlike likelihood-based generative models (e.g., regularized flows \cite{kingma2018glow}), score-based generative models do not require regularization of the generation process. Specifically, in likelihood-based methods, model training usually relies on maximizing the likelihood function, which means that regularization terms need to be introduced to prevent overfitting and ensure model stability. In contrast, SGMs estimate the gradient of the data distribution by optimizing the score function. This approach usually does not require explicit regularization and reduces the complexity of model training. In addition, SGMs only need to focus on estimating the gradient of the data distribution by learning the score function, avoiding the complexity of accurately modeling the entire distribution process. 

\section{Preliminary Information}
\subsection{Score-based Generative Models}
SGMs are efficient generative models that can generate high-quality data and model complex data distribution. A typical SGMs consists of a forward noising and a backward denoising, where the forward noising gradually adds noise to the real data to transform the data from the real distribution to the noise distribution, and the backward denoising starts from the noise sample and gradually removes the noise using the score function to restore the real data sample. Given an input $\mathbf{X}\in \mathbb{R}^d$ and a complicated data distribution $\mathcal{D}$, a forward noising process can be obtained through a stochastic differential equation (SDE) as follows:
\begin{equation}
\small
\mathbf{X}_{0}\sim\mathcal{D},\mathrm{d}\mathbf{X}_{t}={\mathbf{f}}(\mathbf{X}_{t},t)\mathrm{d}t+\sigma_{t}\mathrm{d}\mathbf{B}_{t},~t\in[0,1]
\end{equation}
where $\sigma_{t}$ the diffusion coefficient, $\mathbf{B}$ represents the Brownian motion.

Assuming that $p_t$ is a probability density function, the reverse denoising process can be established through the reversed time SDE as follows:
\begin{equation}
\small
\begin{aligned}
\mathrm{d}\bar{\mathbf{X}}_{t}&=(\mathbf{f}(\bar{\mathbf{X}}_{t},t)-\sigma_{t}^{2}\nabla\log p_{t}(\bar{\mathbf{X}}_{t}))\mathrm{d}\bar{t}+\sigma_{t}\mathrm{d}\bar{\mathbf{B}}_{t}\\ \bar{\mathbf{X}}_{1} & \sim\mathbf{X}_{1},~t\in[0,1]
\end{aligned}
\end{equation}
where $d\bar{t}=-dt$ is the negative infinitesimal time step, $\bar{\mathbf{B}}$ is the reversed time Brownian motion.

In the reverse denoising process, the scoring network $s(\mathbf{X}(t),t;\theta)$ provides gradient information for the current noise sample $\mathbf{X}_t$, indicating how to adjust the value of the sample so as to gradually restore the original data distribution. During the training process, the scoring network is to minimize the gap between the model estimated score function and the true score function through the score matching loss function as follows:
\begin{equation}
\small
\begin{aligned}
\mathcal{L}_{\mathrm{s}}=\mathbb{E}_{t\sim\mathcal{U}(0,T)}[\mathbb{E}_{\mathbf{X}_t|\mathbf{X}_0}[\|s(\mathbf{X}(t),t;\theta)
-\nabla_{\mathbf{X}_t}\log p_{t}(\mathbf{X}_t|X_0)\|_{2}^{2}]
\end{aligned}
\end{equation}
where $\mathcal{U}(0,T)$ is a uniform distribution over $[0, T]$. Given a well-trained scoring network $s_{{\theta}^{*}}$, we can generate realistic data by solving the learned reverse-time SDE as follows:
\begin{equation}
\small
\begin{aligned}
\mathrm{d}\hat{\mathbf{X}}_{t}&=\!(\mathbf{f}(\hat{\mathbf{X}}_{t},t)-\sigma_{t}^{2}s_{{\theta}^{*}}(\hat{\mathbf{X}}_{t}))\mathrm{d}\bar{t}+\sigma_{t}\mathrm{d}\bar{\mathbf{B}}_{t}\\\hat{\mathbf{X}}_{1}& \sim\!\pi,\ t\in[0,1]
\end{aligned}
\end{equation}
where $\pi$ is the prior information of the data.

\subsection{Score-based Graph Generative Models}

In the graph generation model, given a graph $\mathbf{G(X,A)}\in \mathbb{R}^{n\times d}\times \mathbb{R}^{n\times n}$ with $n$ nodes, where $\mathbf{X}\in \mathbb{R}^{n\times d}$ is the feature vectors of each node with dimension $d$ , and $\mathbf{\mathbf{A}}\in \mathbb{R}^{n\times n}$ is the connection relationship between nodes. The goal of the graph generation model is to learn the underlying data distribution of the graph, A standard Graph Diffusion Score-based Model \cite{jo2022score}, \cite{luo2023fast} gradually generates a perturbation graph through a diffusion process and learns the generation process of the graph through a scoring network. Specifically, given each graph sample $(\mathbf{X}, \mathbf{A})$, a forward noising process is obtained through an SDE as follows:
\begin{equation}
\small
\begin{cases}
\mathrm{d}\mathbf{X}_t=\mathbf{f}^X(\mathbf{X}_t,t)\mathrm{d}t+\sigma_{X,t}\mathrm{d}\mathbf{B}_t^X \\
\mathrm{d}\mathbf{A}_t=\mathbf{f}^A(\mathbf{A}_t,t)\mathrm{d}t+\sigma_{A,t}\mathrm{d}\mathbf{B}_t^A & 
\end{cases}\end{equation}
where $\sigma_{X,t}$, and $\sigma_{A,t}$ the diffusion coefficient, $\mathbf{B}_t^A$, and $\mathbf{B}_t^X$ represent the Brownian motion.

Assuming that $p_t$ is a probability density function, the reverse denoising process can be established through the reversed time SDE as follows:
\begin{equation}
\small
\begin{cases}
\mathrm{d}\bar{\mathbf{X}}_t=\left(\mathbf{f}^X(\bar{\mathbf{X}}_t,t)-\sigma_{X,t}^2\nabla_\mathbf{X}\log p_t(\bar{\mathbf{X}}_t,\bar{\mathbf{A}}_t)\right)\mathrm{d}\bar{t}+\sigma_{X,t}\mathrm{d}\bar{\mathbf{B}}_t^X \\
\mathrm{d}\bar{\mathbf{A}}_t=\left(\mathbf{f}^A(\bar{\mathbf{A}}_t,t)-\sigma_{A,t}^2\nabla_\mathbf{A}\log p_t(\bar{\mathbf{X}}_t,\bar{\mathbf{A}}_t)\right)\mathrm{d}\bar{t}+\sigma_{A,t}\mathrm{d}\bar{\mathbf{B}}_t^A & 
\end{cases}\end{equation}
where $\bar{\mathbf{B}}_t^A$, and $\bar{\mathbf{B}}_t^X$ represent the reversed time Brownian motion.

Given $\mathbf{G}_0$, the joint probability distribution of $\mathbf{X}_t$ and $\mathbf{A}_t$ can be simplified to the product of two simpler distributions \cite{jo2022score}, so that the objective function of denoising score matching can be simplified in form as follows:
\begin{equation}
\small
\begin{aligned}
 & \mathcal{L}_{s}^{{\theta}}=\mathbb{E}_{t\sim\mathcal{U}(0,T)}\mathbb{E}_{\mathbf{G}_t|\mathbf{G}}\|s_{\boldsymbol{\theta}}-\nabla\log p_{t|0}(\mathbf{X}_t|\mathbf{X}_0)\|^2 \\
 & \mathcal{L}_{s}^{{\phi}}=\mathbb{E}_{t \sim\mathcal{U}(0,T)}\mathbb{E}_{\mathbf{G}_t|\mathbf{G}}\|s_{\boldsymbol{\phi}}-\nabla\log p_{t|0}(\mathbf{A}_t|\mathbf{A}_0)\|^2
\end{aligned}\end{equation}

\section{Problem Definition}
Assume that the set $\{\mathbf{X}_1,\mathbf{X}_2,\ldots ,\mathbf{X}_M\}$ represents the $M$ modalities, where $\mathbf{X}_k$  represents the input of the $k$-th modality. We introduce a binary indicator $\alpha \in \{0,1\}$ to identify the availability status of each modality. If the $k$-th modality is missing, let $\alpha_k =0$; conversely, if the $k$-th modality is available, let $\alpha_k=1$. We can define a set of missing modalities $\mathcal{I}_m =\{k|\alpha_k=0\}$. In this incomplete modality scenario, the goal is to recover these unobserved modalities to make up for the missing information. The process of modal recovery usually needs to rely on the existing observed modal information $\mathcal{I}_o=\{k|\alpha_k =1\}$, and complete it by modeling the correlation between modalities.

Our main idea is to recover the missing emotion modality $\mathcal{I}_m$  from its latent distribution space conditioned on the observed modality $\mathcal{I}_\mathrm{o}$. We use the observed modality $\mathcal{I}_\mathrm{o}$  as a semantic condition to guide the generation of the missing modality, ensuring that the recovered modality data is consistent and relevant to the real data. Formally, we denote the data distribution of the missing modality as $p(\mathbf{X}_m)$ and the data distribution of the available modality as $p(\mathbf{X}_{\mathcal{I}_\mathrm{o}})$. Our ultimate goal is to sample the missing modality data from the conditional distribution $p(\mathbf{X}_m|\mathbf{X}_{\mathcal{I}_\mathrm{o}})$. Inspired by graph completion networks \cite{lian2023gcnet} and diffusion modality generation \cite{wang2024incomplete}, we combine the advantages of GNNs and diffusion models to simultaneously model the complementary semantic information between modalities and reconstruct high-quality missing modality features. 

However, directly adding Gaussian noise to the adjacency matrix may seriously destroy the local structure of the graph, resulting in an unreasonable generated adjacency matrix. To overcome these problems, we propose to strictly restrict the diffusion of Gaussian noise to the spectral space of the adjacency matrix. On the one hand, by operating the eigenvalues in the spectral space, the direct destruction of the local structure of the adjacency matrix can be effectively avoided, ensuring that the generated graph still conforms to the global semantics. On the other hand, this constrained diffusion process can more naturally capture and preserve the topological information of the graph, ensuring that the generated graph has good connectivity and consistent spectral features.

Specifically, we consider a multi-step diffusion model to gradually construct the conditional distribution by perturbing $\mathbf{X}_m$  and $\mathbf{\Lambda}_m$, where $\mathbf{A}_m=\mathbf{U}_m\mathbf{\Lambda}_m\mathbf{U}_m$, $\mathbf{U}$ are the eigenvectors and $\mathbf{\Lambda}$ is the diagonal eigenvalues. In the $t$-th step, the conditional transfer distribution of the modal features and the adjacency matrix can be expressed as $p_t(\mathbf{X}_m (t)|\mathbf{X}_{\mathcal{I}_\mathrm{o}}(0))$ and $p_t(\mathbf{\Lambda}_m (t)|\mathbf{\Lambda}_{\mathcal{I}_\mathrm{o}}(0))$ and can be approximated as follows:
\begin{equation}
\small
\begin{aligned}
&p_{t}(\mathbf{X}_{\mathrm{m}}(t)|\mathbf{X}_{\mathcal{I}_{\mathrm{o}}}(0))\\&=\int p_{t}(\mathbf{X}_{\mathrm{m}}(t)|\mathbf{X}_{\mathcal{I}_{\mathrm{o}}}(t),\mathbf{X}_{\mathcal{I}_{\mathrm{o}}}(0))p_{t}(\mathbf{X}_{\mathcal{I}_{\mathrm{o}}}(t)|\mathbf{X}_{\mathcal{I}_{\mathrm{o}}}(0))\mathrm{d}\mathbf{X}_{\mathcal{I}_{\mathrm{o}}}(t) \\ &\approx\int p_{t}(\mathbf{X}_{\mathrm{m}}(t)|\mathbf{X}_{\mathcal{I}_{\mathrm{o}}}(t))p_{t}(\mathbf{X}_{\mathcal{I}_{\mathrm{o}}}(t)|\mathbf{X}_{\mathcal{I}_{\mathrm{o}}}(0))\mathrm{d}\mathbf{X}_{\mathcal{I}_{\mathrm{o}}}(t) \\
&p_{t}(\mathbf{\Lambda}_{\mathrm{m}}(t)|\mathbf{\Lambda}_{\mathcal{I}_{\mathrm{o}}}(0))\\&=\int p_{t}(\mathbf{\Lambda}_{\mathrm{m}}(t)|\mathbf{\Lambda}_{\mathcal{I}_{\mathrm{o}}}(t),\mathbf{\Lambda}_{\mathcal{I}_{\mathrm{o}}}(0))p_{t}(\mathbf{\Lambda}_{\mathcal{I}_{\mathrm{o}}}(t)|\mathbf{\Lambda}_{\mathcal{I}_{\mathrm{o}}}(0))\mathrm{d}\mathbf{\Lambda}_{\mathcal{I}_{\mathrm{o}}}(t) \\ &\approx\int p_{t}(\mathbf{\Lambda}_{\mathrm{m}}(t)|\mathbf{\Lambda}_{\mathcal{I}_{\mathrm{o}}}(t))p_{t}(\mathbf{\Lambda}_{\mathcal{I}_{\mathrm{o}}}(t)|\mathbf{\Lambda}_{\mathcal{I}_{\mathrm{o}}}(0))\mathrm{d}\mathbf{\Lambda}_{\mathcal{I}_{\mathrm{o}}}(t)
\end{aligned}
\end{equation}

According to the score-based diffusion model \cite{jo2022score}, we calculated the conditional transition probability score $p_t(\mathbf{X}_m (t)|\mathbf{X}_{\mathcal{I}_\mathrm{o}}(0))$ and $p_t(\mathbf{\Lambda}_m (t)|\mathbf{\Lambda}_{\mathcal{I}_\mathrm{o}}(0))$ as follows:
\begin{equation}
\small
\begin{aligned}
&\nabla_{\mathbf{X}_{\mathrm{m}}}\log p_{t}(\mathbf{X}_{\mathrm{m}}(t)|\mathbf{X}_{\mathcal{I}_{\mathrm{o}}}(0)) \\&\approx\nabla_{\mathbf{X}_{\mathrm{m}}}\log\mathbb{E}_{p_{t}(\mathbf{X}_{\mathcal{I}_{\mathrm{o}}}(t)|\mathbf{X}_{\mathcal{I}_{\mathrm{o}}}(0))}[p_{t}(\mathbf{X}_{\mathrm{m}}(t)|\mathbf{X}_{\mathcal{I}_{\mathrm{o}}}(t))] \\&\approx\nabla_{\mathbf{X}_{\mathrm{m}}}\log p_{t}(\mathbf{X}_{\mathrm{m}}(t)|\mathbf{X}_{\mathcal{I}_{\mathrm{o}}}(t)) \\&=\nabla_{\mathbf{X}_{\mathrm{m}}}\log p_{t}([\mathbf{X}_{\mathrm{m}}(t);\mathbf{X}_{\mathcal{I}_{\mathrm{o}}}(t)]) \\
&\nabla_{\mathbf{\Lambda}_{\mathrm{m}}}\log p_{t}(\mathbf{\Lambda}_{\mathrm{m}}(t)|\mathbf{\Lambda}_{\mathcal{I}_{\mathrm{o}}}(0)) \\&\approx\nabla_{\mathbf{\Lambda}_{\mathrm{m}}}\log\mathbb{E}_{p_{t}(\mathbf{\Lambda}_{\mathcal{I}_{\mathrm{o}}}(t)|\mathbf{\Lambda}_{\mathcal{I}_{\mathrm{o}}}(0))}[p_{t}(\mathbf{\Lambda}_{\mathrm{m}}(t)|\mathbf{\Lambda}_{\mathcal{I}_{\mathrm{o}}}(t))] \\&\approx\nabla_{\mathbf{\Lambda}_{\mathrm{m}}}\log p_{t}(\mathbf{\Lambda}_{\mathrm{m}}(t)|\mathbf{\Lambda}_{\mathcal{I}_{\mathrm{o}}}(t)) \\&=\nabla_{\mathbf{\Lambda}_{\mathrm{m}}}\log p_{t}([\mathbf{A}_{\mathrm{m}}(t);\mathbf{\Lambda}_{\mathcal{I}_{\mathrm{o}}}(t)])
\end{aligned}
\end{equation}
where $\mathbf{X}_{\mathcal{I}_{\mathrm{o}}}(t)$ and $\mathbf{\Lambda}_{\mathcal{I}_{\mathrm{o}}}(t)$ is a random sample from $p_{t}(\mathbf{X}_{\mathcal{I}_{\mathrm{o}}}(t))|\mathbf{X}_{\mathcal{I}_{\mathrm{o}}}(0))$ and $p_{t}(\mathbf{\Lambda}_{\mathcal{I}_{\mathrm{o}}}(t))|\mathbf{\Lambda}_{\mathcal{I}_{\mathrm{o}}}(0))$, respectively. Eq. 9 is held because:
\begin{equation}
\small
\begin{aligned}
&\nabla_{\mathbf{X}_{m}}\log p_{t}([\mathbf{X}_{\mathrm{m}}(t);\mathbf{X}_{\mathcal{I}_{\mathrm{o}}}(t)])\\&=\nabla_{\mathbf{X}_{m}}\log p_{t}(\mathbf{X}_{\mathrm{m}}(t)|\mathbf{X}_{\mathcal{I}_{\mathrm{o}}}(t))+\nabla_{\mathbf{X}_{m}}\log p_{t}(\mathbf{X}_{\mathcal{I}_{\mathrm{o}}}(t)) \\&=\nabla_{\mathbf{X}_{m}}\log p_{t}(\mathbf{X}_{\mathrm{m}}(t)|\mathbf{X}_{\mathcal{I}_{\mathrm{o}}}(t))\\
&\nabla_{\mathbf{\Lambda}_{m}}\log p_{t}([\mathbf{\Lambda}_{\mathrm{m}}(t);\mathbf{\Lambda}_{\mathcal{I}_{\mathrm{o}}}(t)])\\&=\nabla_{\mathbf{\Lambda}_{m}}\log p_{t}(\mathbf{\Lambda}_{\mathrm{m}}(t)|\mathbf{\Lambda}_{\mathcal{I}_{\mathrm{o}}}(t))+\nabla_{\mathbf{A}_{m}}\log p_{t}(\mathbf{\Lambda}_{\mathcal{I}_{\mathrm{o}}}(t)) \\&=\nabla_{\mathbf{\Lambda}_{m}}\log p_{t}(\mathbf{\Lambda}_{\mathrm{m}}(t)|\mathbf{\Lambda}_{\mathcal{I}_{\mathrm{o}}}(t))
\end{aligned}
\end{equation}

Finally, we derive the score-matching objective as follows:
\begin{equation}
\small
\begin{aligned}
 & \mathcal{L}_{s}^{{\theta}}=\mathbb{E}_{\mathbf{X}_{\mathcal{I}_{\mathrm{o}},\mathbf{X}_{m},t\sim\mathcal{U}(0,T)}}\mathbb{E}_{\mathbf{G}_t|\mathbf{G}}\|s_{\boldsymbol{\theta}}-\nabla\log p_{t|0}(\mathbf{X}_m(t)|\mathbf{X}_{\mathcal{I}_{\mathrm{o}}}(0))\|^2 \\
 & \mathcal{L}_{s}^{{\phi}}=\mathbb{E}_{\mathbf{\Lambda}_{\mathcal{I}_{\mathrm{o}},\mathbf{\Lambda}_{m},t \sim\mathcal{U}(0,T)}}\mathbb{E}_{\mathbf{G}_t|\mathbf{G}}\|s_{\boldsymbol{\phi}}-\nabla\log p_{t|0}(\mathbf{\Lambda}_m(t)|\mathbf{\Lambda}_{\mathcal{I}_{\mathrm{o}}}(0)\|^2
\end{aligned}\end{equation}

\begin{figure*}
    \centering
    \setlength{\abovecaptionskip}{0.cm}
    \includegraphics[width=1\linewidth]{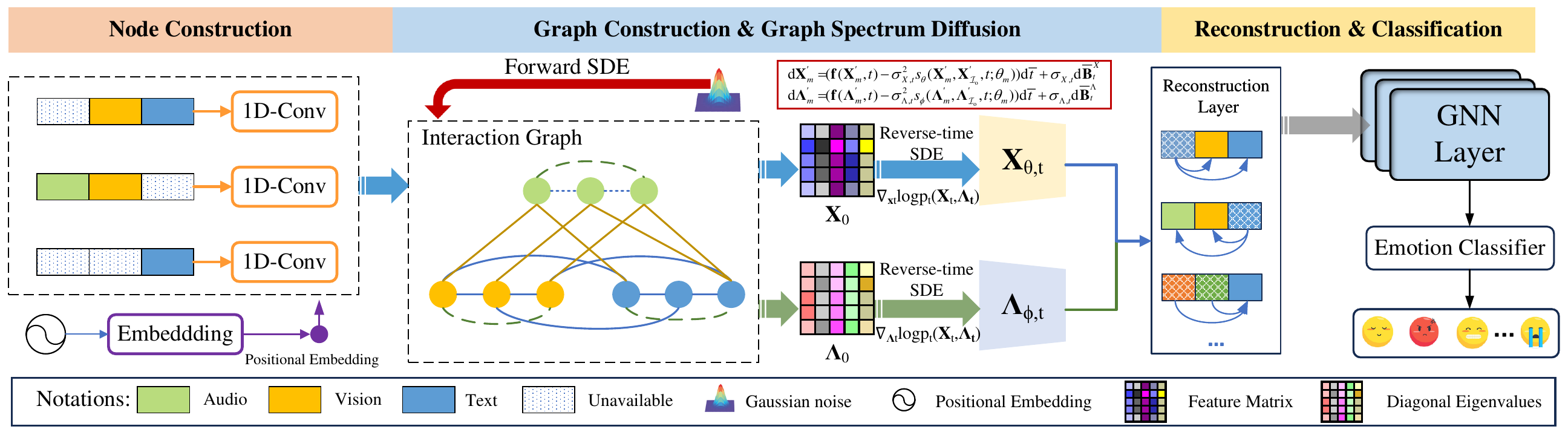}
    \caption{The framework of GSDNet. Given incomplete input data, GSDNet encodes shallow features through 1D-Conv and combines position embedding information. In the missing modal graph diffusion network, we sample from the prior noise distribution and add it to the node features and diagonal eigenvalues, and then solve the inverse time diffusion through the score model to denoise the features to generate new samples. Finally, the reconstructed features are used as complete data to predict the emotion label.}
    \label{fig:enter-label}
    \vspace{-4mm}
\end{figure*}

\section{THE PROPOSED METHOD}

\subsection{Modality Encoder}
Since the original features of text, audio, and video modalities usually have significant dimensional differences, directly using the original features of the modalities to recover missing modalities may lead to difficulties in semantic alignment or even introduce noise.To ensure that the unimodal sequence representations of the three modalities can be mapped to in the same feature space, we input these modalities into a one-dimensional convolutional layer to achieve feature alignment:
\begin{equation}
\small
\begin{aligned}
\mathbf{X}_{\mathrm{m}}^{\prime}=\mathrm{Conv}1\mathbf{D}\left(\mathbf{X}_{\mathrm{m}},l_{m}\right)\in\mathbb{R}^{N\times d},\,m\in\{t,a,v\}
\end{aligned}\end{equation}
where $l_m$ represents the size of the one-dimensional convolution kernel corresponding to the $m$-th modality. $N$ represents the number of utterances in the conversation. $d$ represents the dimension of the common feature space.

To make full use of the position and order information in the sequence, we introduced position embedding when processing the sequence after convolution:
\begin{equation}
\small
\begin{aligned}
{\bf PE}_{(pos,2i)}&=\sin\left(\frac{pos}{10000^{2i/d}}\right)\\
{\bf PE}_{(pos,2i+1)}&=\cos\left(\frac{pos}{10000^{2i/d}}\right)
\end{aligned}\end{equation}
where $pos$ represents the index of the sequence, and dimension $i$ represents the index of the feature dimension.

Overall, we feed position embeddings into a convolutional sequence as follows:
\begin{equation}
\small
\mathbf{X}_{\mathrm{m}}^{\prime}=\mathbf{X}_{\mathrm{m}}^{\prime} + {\bf PE}
\end{equation}

\subsection{Missing Modality Graph Spectral Diffusion Network}
We train two scoring networks $s_\theta$ and $s_\phi$ to model the distribution of missing modalities $m \in \mathcal{I}_m$, respectively. Similar to score-based diffusion \cite{jo2022score}, the corresponding inverse-time SDE can be derived as follows:
\begin{equation}
\small
\begin{aligned}
\mathrm{d}{\mathbf{X}}_{m}^{\prime}&=\!(\mathbf{f}(\mathbf{X}_{m}^{\prime},t)-\sigma_{X,t}^{2}s_{{\theta}}(\mathbf{X}_{m}^{\prime},\mathbf{X}_{\mathcal{I}_{\mathrm{o}}}^{\prime},t;\theta_m))\mathrm{d}\bar{t}+\sigma_{X,t}\mathrm{d}\bar{\mathbf{B}}_{t}^X \\
\mathrm{d}{\mathbf{\Lambda}}_{m}^{\prime}&=\!(\mathbf{f}(\mathbf{\Lambda}_{m}^{\prime},t)-\sigma_{\Lambda,t}^{2}s_{{\phi}}(\mathbf{\Lambda}_{m}^{\prime},\mathbf{\Lambda}_{\mathcal{I}_{\mathrm{o}}}^{\prime},t;\theta_m))\mathrm{d}\bar{t}+\sigma_{\Lambda,t}\mathrm{d}\bar{\mathbf{B}}_{t}^\Lambda
\end{aligned}
\label{eq:15}
\end{equation}

The core of Eq. \ref{eq:15} is to guide the inverse diffusion process through the trained score network, gradually transforming random noise into missing modal data consistent with the true distribution. We assume that the language modality $\mathbf{X}_l$, the visual modality $\mathbf{X}_v$, and the corresponding diagonal eigenvalues $\mathbf{\Lambda}_l$ and $\mathbf{\Lambda}_v$ are observed, while the acoustic modality $\mathbf{X}_a$ and the corresponding diagonal eigenvalues $\mathbf{\Lambda}_a$ are missing. Our goal is to model the missing acoustic modality $\mathbf{X}_a$ and $\mathbf{\Lambda}_a$, and recover their data through the score network $s_\theta$ and $s_\phi$ conditioned on $\mathbf{X}_l$, $\mathbf{X}_v$, $\mathbf{\Lambda}_l$ and $\mathbf{\Lambda}_v$ as follows:
\begin{equation}
\small
\begin{aligned}
\mathbf{X}_{a}^{\prime}(t-\Delta t)&=\mathbf{X}_{a}^{\prime}(t)-\mathbf{f}(\mathbf{X}_{m}^{\prime},t)\\&+\sigma^{2}_{X,t}s_{\theta}(\mathbf{X}_{a}^{\prime}(t),[\mathbf{X}_{l}^{\prime};\mathbf{X}_{v}^{\prime}](t),t;\theta_{a})\Delta t\\
&+\sigma_{X,t}\sqrt{\Delta t}\epsilon_{X,t} \\
\mathbf{\Lambda}_{a}^{\prime}(t-\Delta t)&=\mathbf{\Lambda}_{a}^{\prime}(t)-\mathbf{f}(\mathbf{\Lambda}_{m}^{\prime},t)\\&+\sigma^{2}_{\Lambda,t}s_{\phi}(\mathbf{\Lambda}_{a}^{\prime}(t),[\mathbf{\Lambda}_{l}^{\prime};\mathbf{\Lambda}_{v}^{\prime}](t),t;\theta_{a})\Delta t+\sigma_{\Lambda,t}\sqrt{\Delta t}\epsilon_{t}
\end{aligned}
\end{equation}
where $\Delta t$ is a discrete time step size and $\epsilon_{t} \sim \mathcal{N} (0, I)$. After enough iterations, we can gradually guide the noise data to approach the target distribution and finally obtain the restored acoustic modal data $\mathbf{X}_{a}^{\prime}$ and the corresponding diagonal eigenvalues $\mathbf{\Lambda}_a^{\prime}$. To generate more refined acoustic modalities, we input the restored acoustic data $xa$ into a specially designed acoustic modal reconstruction module $\mathcal{D}_X$ and diagonal eigenvalues reconstruction module $\mathcal{D}_\Lambda$ to obtain the final reconstructed acoustic modal $\hat{\mathbf{X}}_a=\mathcal{D}_X(\mathbf{X}_a^{\prime})$ and diagonal eigenvalues $\hat{\mathbf{\Lambda}}_a=\mathcal{D}_\Lambda(\mathbf{\Lambda}_a^{\prime})$. We define a reconstruction loss function $\mathcal{L}_{rec}$ to measure the difference between the reconstructed data and the original target modal data under any missing patterns as follows:
\begin{equation}
\small
\begin{aligned}
\mathcal{L}_{\text{rec}} = \sum_{i \in \mathcal{I}_{\text{m}}} \left\| \hat{\mathbf{X}}_i - \mathbf{X}_i \right\|_2^2+\left\| \hat{\mathbf{\Lambda}}_i - \mathbf{\Lambda}_i \right\|_2^2
\end{aligned}
\end{equation}

Therefore, the loss of the missing modality graph diffusion network is as follows:
\begin{equation}
\small
\begin{aligned}
\mathcal{L}_{\text{miss}} = \mathcal{L}_{\text{rec}} + \mathcal{L}_{s}^{{\theta}} + \mathcal{L}_{s}^{{\phi}}
\end{aligned}
\end{equation}

\subsection{Multimodal Fusion and Prediction}

The recovered data and the observed available data are combined to obtain the complete multimodal data $\mathbf{H}$ and the adjacency matrix $\mathbf{A}$. To achieve the fusion of multimodal data, we use GCN to capture the complementary semantic information between the modalities as follows:
\begin{equation}
\mathbf{H}^{(l+1)}=\mathbf{ReLU}(\mathbf{A}\mathbf{H}^{(l)}\mathbf{W}^{(l)})
\end{equation}
where $\mathbf{W}^{(l)}$ is the learnable weight matrix of the $l$-th layer.

To train the entire model, we combine the losses of the above reconstruction and prediction tasks into a joint optimization objective function as follows:
\begin{equation}
\small
\begin{aligned}
\mathcal{L}_{\text{total}} = \beta\mathcal{L}_{\text{miss}} + \mathcal{L}_{\text{pred}} 
\end{aligned}
\end{equation}
where $\beta$ is a hyperparameter.

\begin{table*}[htbp]
\centering
\caption{The performance of different methods is shown under different missing modalities on the CMU-MOSI and CMU-MOSEI datasets. The values reported in each cell represent the $\mathrm{ACC_2}$/$\mathrm{F1}$/$\mathrm{ACC_7}$. Bold indicates the best performance.}
\vspace{-3mm}
\label{TABLE:1}
\resizebox{1\textwidth}{!}{
\begin{tabular}{l|c|cccccc}
\toprule
Datasets &Available & MCTN  & MMIN  & GCNet & DiCMoR & IMDer & GSDNet (Ours) \\
\midrule
\multirow{8}{*}{CMU-MOSI} 
& $\{l\}$  & 79.1/79.2/41.0 & 83.8/83.8/41.6 & 83.7/83.6/42.3 & 84.5/84.4/44.3 & {84.8}/{84.7}/{44.8} & \textbf{86.4/86.6/45.7} \\
& $\{v\}$  & 55.0/54.4/16.3 & 57.0/54.0/15.5 & 56.1/55.7/16.9 & 62.2/60.2/20.9 & {61.3}/{60.8}/{22.2} & \textbf{64.1/63.7/25.3}\\
& $\{a\}$  & 56.1/54.5/16.5 & 55.3/51.5/15.5 & 56.1/54.5/16.6 &62.2/60.2/20.9 & {62.0}/{62.2}/{22.0} & \textbf{64.4/64.1/24.6}\\
& $\{l,v\}$  & 81.1/81.2/42.1 & 83.8/83.9/42.0 & 84.3/84.2/43.4 &85.5/85.4/45.2 & {85.5}/{85.4}/{45.3} & \textbf{86.5/86.4/46.7}\\
& $\{l,a\}$  & 81.0/81.0/43.2 & 84.0/84.0/42.3 & 84.5/84.4/43.4 &85.5/85.5/44.6 & {85.4}/{85.3}/{45.0} & \textbf{86.7/86.6/46.8}\\
& $\{v,a\}$  & 57.5/57.4/16.8 & 60.4/58.5/19.5 & 62.0/61.9/17.2 &64.0/63.5/21.9 & {63.6}/{63.4}/{23.8} & \textbf{65.2/64.8/24.9}\\
& $\{l,v,a\}$  & 81.4/81.5/43.4 & 84.6/84.4/44.8 & 85.2/85.1/44.9  & 85.7/85.6/45.3 & {85.7}/{85.6}/{45.3} & \textbf{87.7/87.3/46.8}\\
& Average & 70.2/69.9/31.3 & 72.7/71.4/31.6 & 73.1/72.8/32.1 & 75.4/75.1/34.7 &  {75.5}/{75.3}/{35.5} & \textbf{77.3/77.1/37.3}\\
\midrule
\midrule
\multirow{8}{*}{CMU-MOSEI} 
& $\{l\}$ & 82.6/82.8/50.2 & 82.3/82.4/51.4 & 83.0/83.2/51.2 & 84.2/84.3/52.4 & {84.5}/{84.5}/{52.5} & \textbf{86.6/86.1/55.3}\\
& $\{v\}$ & 62.6/57.1/41.6 & 59.3/60.0/40.7 & 61.9/61.6/41.7 & 63.6/63.6/42.0 & {63.9}/{63.6}/{42.6} & \textbf{65.1/65.7/44.9}\\
& $\{a\}$ & 62.7/54.5/41.4 & 58.9/59.5/40.4 & 60.2/60.3/41.1 & 62.9/60.4/41.4 & {63.8}/{60.6}/{41.7} & \textbf{64.6/64.2/43.1}\\
& $\{l,v\}$ & 83.2/83.2/50.4 & 83.8/83.4/51.2 & 84.3/84.4/51.1 & 84.9/84.9/53.0 & {85.0}/{85.0}/{53.1} & \textbf{87.3/87.0/56.2}\\
& $\{l,a\}$ & 83.5/83.3/50.7 & 83.7/83.3/52.0 & 84.3/84.4/51.3 & 85.0/84.9/52.7 & {85.1}/{85.1}/{53.1} & \textbf{86.2/86.4/55.5}\\
& $\{v,a\}$ & 63.7/62.7/42.1 & 63.5/61.9/41.8 & 64.1/57.2/42.0 & 65.2/64.4/42.4 & {64.9}/{63.5}/{42.8} & \textbf{66.7/66.3/45.2}\\
& $\{l,v,a\}$ & 84.2/84.2/51.2 & 84.3/84.2/52.4 & 85.2/85.1/51.5 & 85.1/85.1/53.4 & {85.1}/{85.1}/{53.4} & \textbf{87.3/87.2/54.9}\\
& Average & 74.6/72.5/46.8 & 73.7/73.5/47.1 & 74.7/73.7/47.1 & 75.8/75.4/48.2 & {76.0}/{75.3}/{48.5} & \textbf{77.7/77.6/50.7}\\
\bottomrule
\end{tabular}}
\vspace{-3mm}
\end{table*}

\section{EXPERIMENTAL DATABASES AND SETUP}

\subsection{Datasets}

We conduct extensive experiments on two MERC datasets to conduct experiments, including CMU-MOSI \cite{zadeh2016multimodal}, and CMU-MOSEI \cite{zadeh2018multimodal}. On the two datasets, we extract the lexical modality features via pre-trained RoBERTa-Large model \cite{liu2019roberta} and obtain a 1024-dimensional word embedding. For visual modality, each video frame was encoded via DenseNet model \cite{huang2017densely} and obtain a 1024-dimensional visual feature. The acoustic modality was processed by wav2vec \cite{schneider2019wav2vec} to obtain the 512-dimensional acoustic features.




\begin{table*}[ht]
\centering
\caption{The performance of different methods is shown at different missing ratios on the CMU-MOSI and CMU-MOSEI datasets. The values reported in each cell represent the $\mathrm{ACC_2}$/$\mathrm{F1}$/$\mathrm{ACC_7}$. Bold indicates the best performance.}
\label{TABLE:2}
\vspace{-3mm}
\resizebox{1\textwidth}{!}{
\begin{tabular}{l|c|cccccc}
\toprule
Datasets & Missing Rate & MCTN & MMIN & GCNet & DiCMoR & IMDer & GSDNet (Ours) \\
\midrule
\multirow{9}{*}{CMU-MOSI} 
& 0.0  & 81.4/81.5/43.4 & 84.6/84.4/44.8 & 85.2/85.1/44.9 & 85.7/85.6/45.3 & 85.7/85.6/45.3 & \textbf{87.7/87.3/46.8}\\
& 0.1  & 78.4/78.5/39.8 & 81.8/81.8/41.2 & 82.3/82.3/42.1 & 83.9/83.9/43.6 & 84.9/84.8/44.8 & \textbf{87.1/86.5/46.2}\\
& 0.2  & 75.6/75.7/38.5 & 79.0/79.1/38.9 & 79.4/79.5/40.0 & 83.9/83.9/43.6 & 83.5/83.4/44.3 & \textbf{86.4/86.1/45.2}\\
& 0.3  & 71.3/71.2/35.5 & 76.1/76.2/36.9 & 77.2/77.2/38.2 & 80.4/80.2/40.6 & 81.2/81.0/42.5 & \textbf{85.2/85.0/44.3}\\
& 0.4  & 68.0/67.6/32.9 & 71.7/71.6/34.9 & 74.3/74.4/36.6 & 77.9/77.7/37.6 & 78.6/78.5/39.7 & \textbf{83.3/82.9/42.1}\\
& 0.5  & 65.4/64.8/31.2 & 67.2/66.5/32.2 & 70.0/69.8/33.9 & 76.7/76.4/36.4 & 76.2/75.9/37.9 & \textbf{81.2/81.1/40.6}\\
& 0.6  & 63.8/62.5/29.7 & 64.9/64.0/29.1 & 67.7/66.7/29.8 & 73.3/73.0/32.7 & 74.7/74.0/35.8 & \textbf{80.1/79.7/38.7}\\
& 0.7  & 61.2/59.0/27.5 & 62.8/61.0/28.4 & 65.7/65.4/28.1 & 71.1/70.8/30.0 & 71.9/71.2/33.4 & \textbf{77.6/77.3/35.6}\\
& Average  & 70.6/70.1/34.8 & 73.5/73.1/35.8 & 75.2/75.1/36.7 & 78.9/78.7/38.5 & 79.6/79.3/40.5 & \textbf{83.6/83.2/42.3}\\
\midrule
\midrule
\multirow{9}{*}{CMU-MOSEI} 
& 0.0  & 84.2/84.2/51.2 & 84.3/84.2/52.4 & 85.2/85.1/51.5 & 78.9/78.7/38.5 & 85.1/85.1/53.4 & \textbf{87.3/87.2/54.9} \\
& 0.1  & 81.8/81.6/49.8 & 81.9/81.3/50.6 & 82.3/82.1/51.2 & 78.9/78.7/38.5 & 84.8/84.6/53.1 & \textbf{86.7/86.5/54.2}\\
& 0.2  & 79.0/78.7/48.6 & 79.8/78.8/49.6 & 80.3/79.9/50.2 & 81.8/81.5/51.4 & 82.7/82.4/52.0 & \textbf{85.3/85.1/53.5}\\
& 0.3  & 76.9/76.2/47.4 & 77.2/75.5/48.1 & 77.5/76.8/49.2 & 79.8/79.3/50.3 & 81.3/80.7/51.3 & \textbf{83.3/83.0/52.2}\\
& 0.4  & 74.3/74.1/45.6 & 75.2/72.6/47.5 & 76.0/74.9/48.0 & 78.7/77.4/48.8 & 79.3/78.1/50.0 & \textbf{81.4/81.2/51.4}\\
& 0.5  & 73.6/72.6/45.1 & 73.9/70.7/46.7 & 74.9/73.2/46.7 & 77.7/75.8/47.7 & 79.0/77.4/49.2 & \textbf{80.5/80.1/50.7}\\
& 0.6  & 73.2/71.1/43.8 & 73.2/70.3/45.6 & 74.1/72.1/45.1 & 77.7/75.8/47.7 & 78.0/75.5/48.5 & \textbf{79.4/79.1/49.4}\\
& 0.7  & 72.7/70.5/43.6 & 73.1/69.5/44.8 & 73.2/70.4/44.5 & 75.4/72.2/46.2 & 77.3/74.6/47.6 & \textbf{78.2/78.1/48.6}\\
& Average  & 77.0/76.1/46.9 & 77.3/75.4/48.2 & 77.9/76.8/48.3 & 79.9/78.6/49.6 & 80.9/79.8/50.6 & \textbf{82.8/82.5/51.9}\\
\bottomrule
\end{tabular}}
\vspace{-4mm}
\end{table*}

\subsection{Baselines}
We compare our proposed method GSDNet to the state-of-the-art incomplete learning methods, including MCTN, MMIN \cite{zhao2021missing}, GCNet \cite{lian2023gcnet}, DiCMoR \cite{wang2023distribution}, IMDer \cite{wang2023incomplete}.

\section{RESULTS AND DISCUSSION}

\subsection{Comparison with the state-of-the-arts}

Tables \ref{TABLE:1} and \ref{TABLE:2} lists the quantitative results of the different missing modalities and the random missing ratio on CMU-MOSI and CMU-CMSEI datasets, showing the performance of different methods under the missing modal. Specifically, GSDNet achieved the best results on the two datasets, verifying its superiority in dealing with modal missing. The performance improvement of GSDNet may be attributed to its ability to explicitly restore the missing modality, which not only helps to restore the lost information but also provides additional complementary information for MERC. In addition, GSDNet has a significant advantage in maintaining consistency between the restored modality and the original modality. This distribution consistency ensures that the information fusion between different modalities is smoother and more accurate, further improving the overall performance of the model. Compared with other MERC methods, the performance degradation of GSDNet decreases as the modal missing rate increases. In practical applications, when the modal missing rate is high, most recovery-based models will experience significant performance degradation.


\subsection{Ablation study}
We conduct ablation experiments on the CMU-MOSI and CMU-MOSEI datasets. The results in Table \ref{tab:aba} show that GSDNet consistently outperforms all variants. Removing the frequency diffusion degrades the performance, which highlights the role of frequency diffusion in capturing the distribution of multimodal data.

\begin{table}[h]
\centering
\renewcommand\tabcolsep{2.5pt}
\renewcommand\arraystretch{0.4}
\caption{Ablation study of graph spectral diffusion on GSDNet under average random missing ratios.}
\vspace{-3mm}
\begin{tabular}{lccc|ccc}
\toprule
\multirow{2}{*}{Methods} & \multicolumn{3}{c|}{CMU-MOSI} & \multicolumn{3}{c}{CMU-MOSEI} \\
\cmidrule(lr){2-4} \cmidrule(lr){5-7}
& ACC$_2$ & F1 & ACC$_7$ & ACC$_2$ & F1 & ACC$_7$ \\
\midrule
GSDNet & 75.7 & 70.6 & 35.3 & 78.1 & 77.4 & 47.4 \\
GSDNet w/spectral & \textbf{83.6} & \textbf{83.2} & \textbf{42.3} & \textbf{82.8} & \textbf{82.5} & \textbf{51.9} \\
\bottomrule
\end{tabular}
\label{tab:aba}
\vspace{-3mm}
\end{table}

\begin{figure*}
    \centering
    \setlength{\abovecaptionskip}{0.cm}
    \includegraphics[width=1\linewidth]{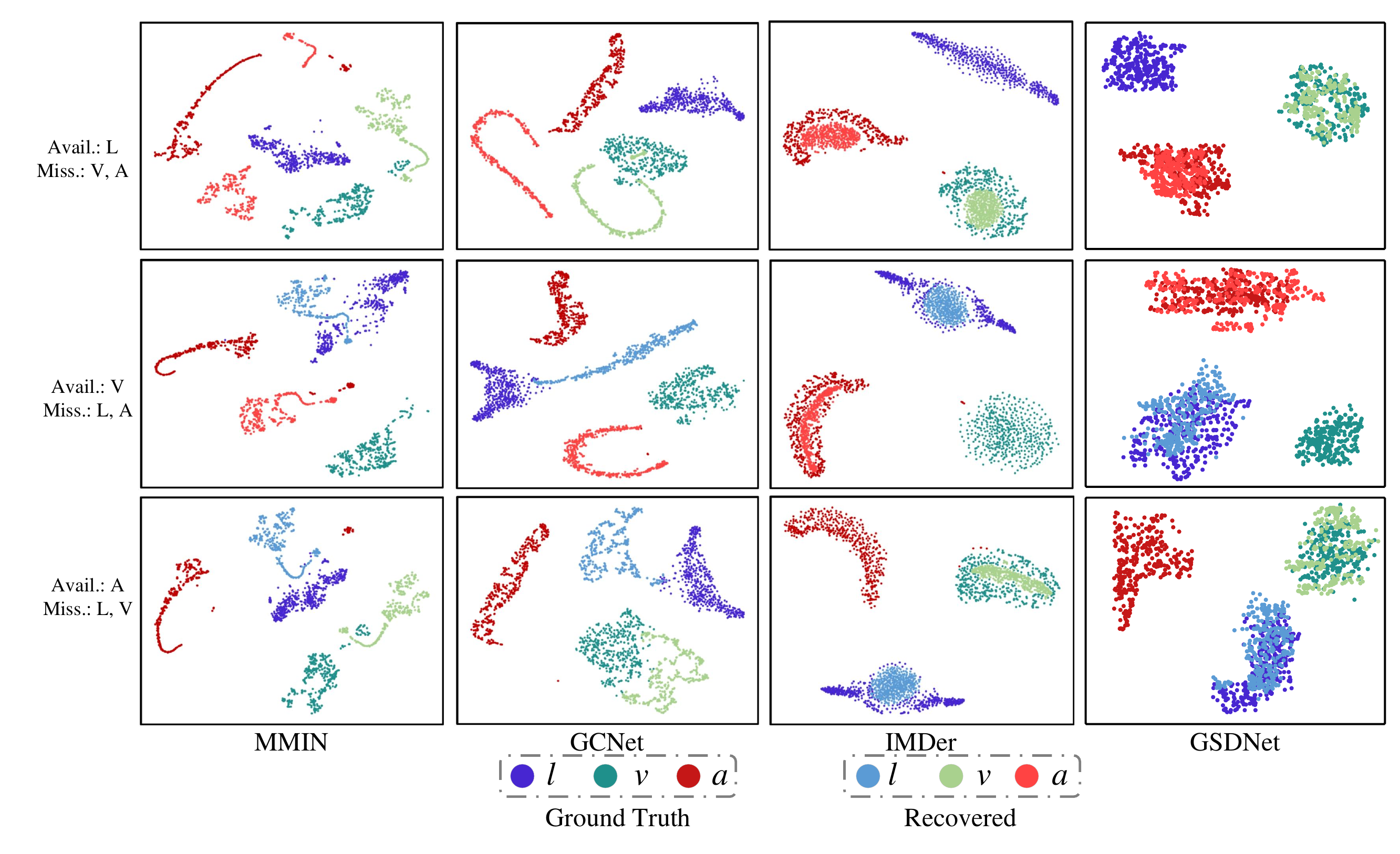}
    \caption{Visualization of restored modalities. Avail. indicates available.}
    \label{fig:vis}
    \vspace{-3mm}
\end{figure*}

\subsection{Visualization of Embedding Space}
Fig. \ref{fig:vis} shows the distribution of the restored data and the original data in the feature space obtained by different restoration methods under the condition of fixed missing modalities. In order to compare these distributions more intuitively, we use t-SNE dimensionality reduction technology on the CMU-MOSEI dataset to project the high-dimensional features into two-dimensional space for visualization. As can be seen from Fig. \ref{fig:vis}, the modal data restored by GSDNet is closest to the distribution of the original data, which shows that GSDNet can better maintain the original feature distribution of the data when restoring the missing modalities. In contrast, there is a clear difference between the distribution of the restored data of other methods and the original data, especially in some local areas, the degree of overlap of the distribution is low.

\begin{figure}[htbp]
	\centering
    \setlength{\abovecaptionskip}{0.cm}
	\subfloat[CMU-MOSI]{\includegraphics[width=0.495\linewidth]{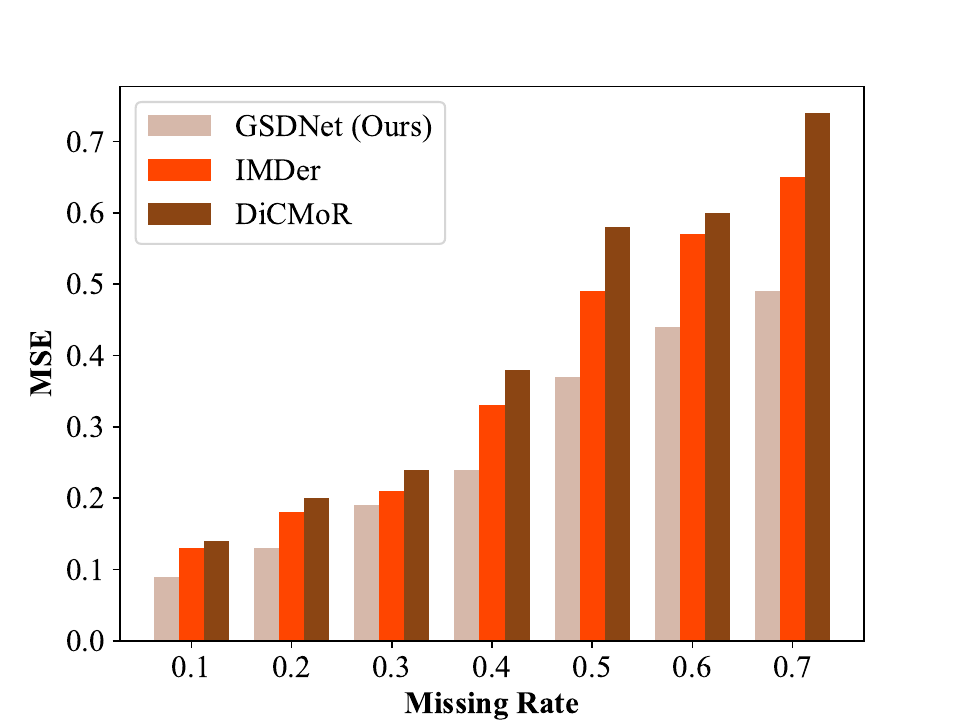}
		\label{fig:mosi}}
	\hfil
	\subfloat[CMU-MOSEI]{\includegraphics[width=0.48\linewidth]{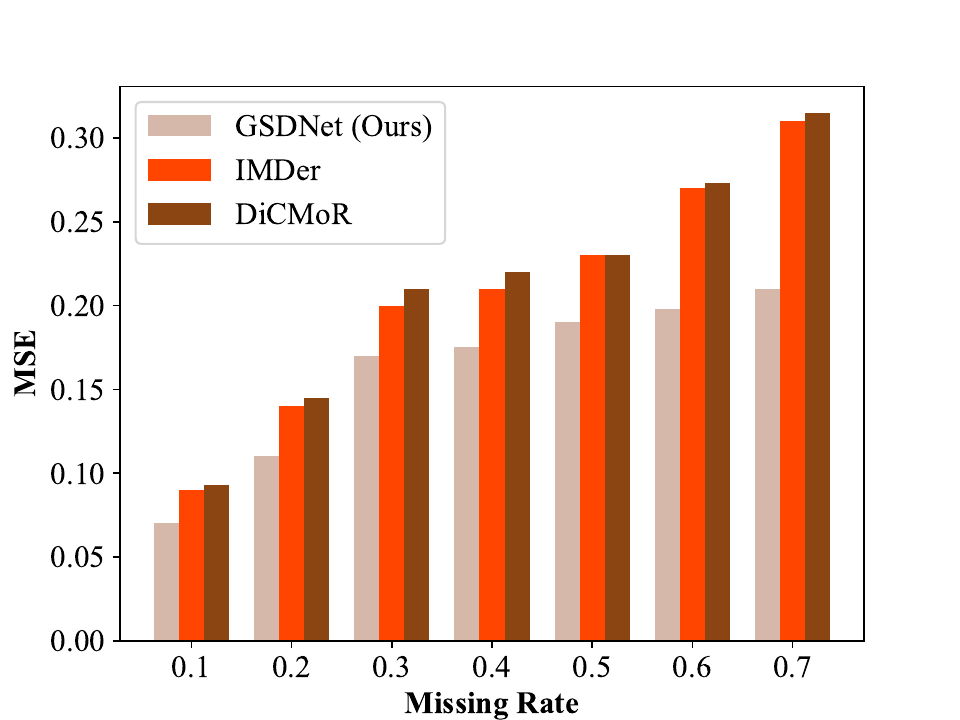}
		\label{fig:mosei}}
    \caption{The comparison of interpolation performance under different missing rates shows the interpolation effects of various methods when dealing with different missing rates.}
	\label{fig.1}
    \vspace{-3mm}
\end{figure}

\subsection{Imputation Performance}
Fig. \ref{fig.1} shows the interpolation results of different methods under different missing rates. By comparing the performance of baseline models under different missing rates, our proposed GSDNet always outperforms other baseline methods in the CMU-MOSI and CMU-MOSEI datasets and all missing rate conditions. Specifically, GSDNet not only shows strong interpolation performance under low missing rates but also has more outstanding performance advantages under high missing rates. The experimental results show that speaker dependency and data distribution consistency play a vital role in data interpolation tasks. Most baseline methods often ignore the synergy of these dependencies, which limits their interpolation performance when dealing with missing data. In contrast, GSDNet can use the speakers relationship to perform more accurate interpolation while maintaining data distribution consistency through the graph diffusion model, so that GSDNet can always maintain relatively good performance under various missing rates.

\section{Conclusions}

In this paper, we introduce a novel Graph Spectral Diffusion Network (GSDNet), which maps Gaussian noise to the graph spectral space of missing modalities and recover the missing data according to original distribution. Compared with previous graph diffusion methods, GSDNet only affects the eigenvalues of the adjacency matrix instead of destroying the adjacency matrix directly, which can maintain the global topological information and important spectral features during the diffusion process. Extensive experiments have demonstrated that GSDNet achieves state-of-the-art emotion recognition performance in various modality loss scenarios.

\section*{Acknowledgments}\label{sec11}
This work is supported by the National Natural Science Foundation of China (Grant No. 62372478), the Research Foundation of Education Bureau of Hunan Province of China (Grant No. 22B0275), and the Hunan Provincial Natural Science Foundation Youth Project (Grant No. 2025JJ60420).

\bibliographystyle{named}
\bibliography{refs}

\end{document}